\documentclass[12pt]{article}
\usepackage{a4wide}
\usepackage{latexsym}
\usepackage{amsmath}
\usepackage{amsfonts}
\usepackage{amscd}
\usepackage{amssymb}
\usepackage{cite}
\usepackage{graphicx}
\usepackage{float}

\usepackage{pslatex}
\usepackage[latin1]{inputenc}
\usepackage[OT2,T1]{fontenc}

\newcommand{\bq}{\begin{eqnarray}}
\newcommand{\eq}{\end{eqnarray}}

\DeclareSymbolFont{cyrletters}{OT2}{wncyr}{m}{n}
\DeclareMathSymbol{\Sha}{\mathalpha}{cyrletters}{"58}

\usepackage{color}

\begin{document}

\thispagestyle{empty}

\begin{flushright}
  ANL-HEP-PR-12-82 \\
  MZ-TH/12-47
% \\ version of \today
\end{flushright}

\vspace{1.5cm}

\begin{center}
  {\Large\bf Direct numerical integration for multi-loop integrals\\
  }
  \vspace{1cm}
  {\large Sebastian Becker ${}^{a}$ and Stefan Weinzierl ${}^{b}$\\
  \vspace{1cm}
      {\small ${}^{a}$ \em High Energy Physics Division}\\
      {\small \em Argonne National Laboratory, Argonne, IL 60439, USA}\\
  \vspace{2mm}
      {\small ${}^{b}$ \em PRISMA Cluster of Excellence, Institut f{\"u}r Physik, }\\
      {\small \em Johannes Gutenberg-Universit{\"a}t Mainz,}\\
      {\small \em D - 55099 Mainz, Germany}\\
  } 
\end{center}

\vspace{2cm}

% abstract ---------------------------------------
\begin{abstract}\noindent
  {
We present a method to construct a suitable contour deformation in loop momentum space for multi-loop integrals.
This contour deformation can be used to perform the integration for multi-loop integrals numerically.
The integration can be performed directly in loop momentum space without the introduction of Feynman or Schwinger parameters.
The method can be applied to finite multi-loop integrals and to divergent multi-loop integrals with suitable subtraction terms.
The algorithm extends techniques from the one-loop case to the multi-loop case.
Examples at two and three loops are discussed explicitly.
   }
\end{abstract}

\vspace*{\fill}

% main text ------------------------------------
\newpage

\section{Introduction}

The automation of NNLO calculations is an ambitious project.
Among other methods, a purely numerical method can be one approach to achieve this goal.
To be more specific we think about extending the numerical method used 
at NLO \cite{Becker:2012nk,Gotz:2012zz,Becker:2012aq,Becker:2011vg,Becker:2010ng,Assadsolimani:2010ka,Assadsolimani:2009cz,Gong:2008ww,Nagy:2003qn}
towards the NNLO case.
The main ingredients within this approach would be a set of subtraction terms to render all contributions individually finite and a method
for the contour deformation such that loop integrals can be performed numerically by Monte Carlo methods.
For performance reasons this numerical loop integration will then be combined with the integration over the phase space of the final particles 
in a single Monte Carlo integration.

As a first step in this direction we consider in this paper a method for the contour deformation of multi-loop integrals.
We generalise the method for the contour deformation from the one-loop case to the multi-loop case and put special emphasis on the cases
of two and three loops, which are the most interesting cases for applications.

As in the one-loop case there are several variants on how the contour deformation can be performed.
The main difference is given by the fact that some variants use additional Feynman parameters while others do not.
The use of Feynman parameters has the advantage that the construction of the deformation is relatively straightforward.
This method has already been discussed in the literature for the multi-loop case \cite{Borowka:2012yc,Kurihara:2005ja,Anastasiou:2007qb,Nagy:2006xy,Yuasa:2011ff,deDoncker:2004fb}.
On the other hand, Feynman parametrisation combines a product of $n$ propagators into a single term, 
which is raised to the power $n$.
The power $n$ magnifies the statistical Monte Carlo integration error we get from regions with integrable singularities.
This can be acceptable for small values of $n$, 
but for multi-parton final states it is better to develop a method which avoids this problem from the start. 
This can be achieved by deforming the contour directly in loop momentum space
without the introduction of Feynman parameters.
However, in this case the construction of the deformation is more challenging.
In this paper we present an algorithm which constructs for a multi-loop integral the deformation vector in loop momentum space.
The algorithm builds upon an algorithm for the one-loop case and considers all possible cycles (i.e. one-loop sub-diagrams) within the multi-loop diagram.
We show that a naive iterative procedure (e.g. picking a one-loop sub-diagram and treating the remainder recursively as a $(l-1)$-loop diagram)
will not work. The sum over all cycles is essential.

The paper is organised as follows:
In the next section we introduce our notation and the basic concepts.
The algorithm for the direct contour deformation in the loop momentum space for multi-loop integrals is presented in section~\ref{sect:algorithm}.
Section~\ref{sect:checks} provides checks and examples.
Our conclusions are given in section~\ref{sect:conclusions}.
The algorithm for the multi-loop case builds upon a method for the one-loop case.
A method for the contour deformation in the one-loop case is reviewed in appendix~\ref{sect:one_loop}.

% ----------------------------------------------------------------------------------
\section{Notation}
\label{sect:notation}

In this section we define the notation.
We consider a $l$-loop Feynman graph with $m$ external lines and $n$ internal lines.
We label the external momenta by $p_1$, ..., $p_m$ and the $l$ independent loop momenta by
$k_1$, ..., $k_l$.
The momenta flowing through the internal lines are denoted by $q_i$ with $1 \le i \le n$.
The momenta $q_i$ can be expressed
as a linear combination of the external 
momenta $p_j$ and the loop momenta $k_j$ with coefficients $-1$, $0$ or $1$:
\bq
\label{eq_internal_mom}
 q_i & = & \sum\limits_{j=1}^l \rho_{ij} k_j + \sum\limits_{j=1}^m \sigma_{ij} p_j,
 \;\;\;\;\;\; 
 \rho_{ij}, \sigma_{ij} \in \{-1,0,1\}.
\eq
We denote an internal propagator by
\bq
\label{def_propagator}
 D_{i} & = & \frac{1}{q_{i}^{2}-m_{i}^{2}+i0}.
\eq
The infinitesimal imaginary part $i0$ indicates the direction into which the contour should be deformed 
in the case where the propagator goes on-shell and is not pinched.
The integral which we want to consider is given by
\bq
\label{eq_basic_integral}
I  & = &
 \int \prod\limits_{r=1}^{l} \frac{d^4k_r}{(2\pi)^4}
 \;
 R(k_1,...,k_l)
 \prod\limits_{j=1}^{n} D_j.
\eq
We assume that the integral is finite. 
In other words, we assume that the function $R(k_1,...,k_l)$ contains appropriate subtraction terms such that the integral is finite.
The function $R(k_1,...,k_l)$ is either a polynomial in the loop momenta $k_1$, ..., $k_l$, or -- more generally -- a
rational function in the loop momenta with poles which are sufficiently far away from the integration contour.
The latter possibility occurs already at one-loop through the ultraviolet subtraction terms, which introduce a new ``ultraviolet'' propagator
with an arbitrary mass $\mu_{\mathrm{UV}}$. 
Beyond one-loop there will be more than one ultraviolet propagator.
Taking $\mu_{\mathrm{UV}}^2$ large enough on the negative imaginary axis avoids that the integration contour
comes close to this pole.
We consider therefore the poles of $R(k_1,...,k_l)$ to be harmless and seek an integration contour, which avoids whenever possible the poles of the
propagators $D_j$.
The integration contour is entirely determined by the $n$ propagators $D_j$.

\begin{figure}
\begin{center}
\includegraphics[width=0.45\textwidth]{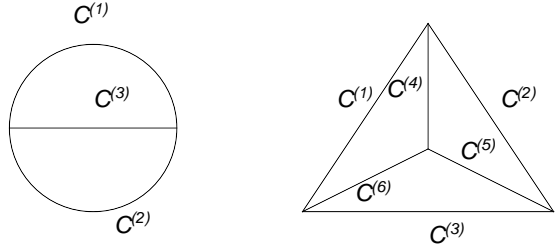}
\caption{\label{fig_chain} 
The generic chain diagrams at two-loop (left) and three-loop (right).
The two-loop chain diagram consists of three chains, the three-loop chain diagram consists of six chains.
}
\end{center}
\end{figure}

It will be useful to group the internal propagators $D_j$ into chains \cite{Kinoshita:1962ur}.
\begin{figure}
\begin{center}
\includegraphics[viewport=140 535 460 650]{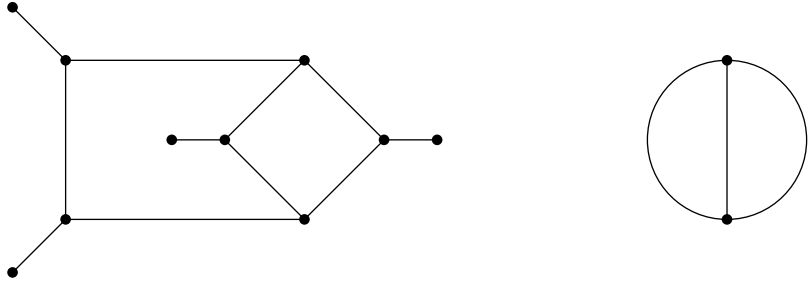}
\end{center}
\caption{
The graph of the crossed double box (left) and the associated chain diagram (right). 
}
\label{fig_crossedbox}
\end{figure}
Two propagators belong to the same chain, if their momenta differ only by a linear combination of the external momenta.
In the representation of eq.~(\ref{eq_internal_mom}) two internal propagators belong to the same chain if and only if the corresponding
matrices $\rho_{ij}$ are identical up to an overall sign.
Obviously, each internal line can only belong to one chain. 
We denote the number of different chains by $c$.
From a given Feynman graph we obtain a new graph which we call the chain diagram by deleting all external lines and by choosing one propagator for each chain
as a representative.
We illustrate this with a few examples.
In fig.~(\ref{fig_crossedbox}) we show the two-loop graph of the crossed double box
and the associated chain diagram which is obtained by deleting the four external lines.
There are three chains, containing three, two and two propagators, respectively.
For each chain we only draw one propagator as a representative, resulting in the chain diagram shown in fig.~(\ref{fig_crossedbox}) on the right.
\begin{figure}
\begin{center}
\includegraphics[viewport=140 535 460 650]{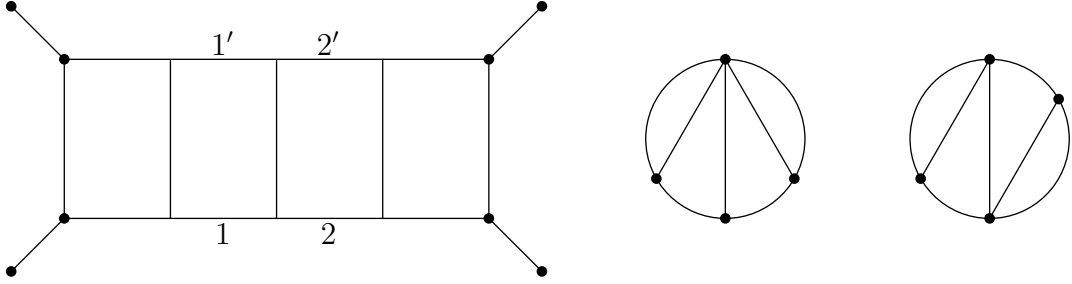}
\end{center}
\caption{
The quadruple box (left) and two possible chain diagrams (middle and right).
The propagators labelled $1$ and $1'$ belong to the same chain, as do the propagators labelled $2$ and $2'$. 
}
\label{fig_quadruplebox}
\end{figure}
Since we choose a propagator as a representative for each chain, the associated chain diagram for a given Feynman graph is not unique.
This is illustrated in fig.~(\ref{fig_quadruplebox}) for the quadruple box.
In the graph of the quadruple box the two propagators labelled $1$ and $1'$ belong to the same chain, as do the propagators labelled $2$ and $2'$.
If we choose the propagators $1$ and $2$ as representatives for the two chains, we obtain the chain diagram shown in the 
middle in fig.~(\ref{fig_quadruplebox}).
If on the other hand we choose $1$ and $2'$ as representatives, we obtain the chain diagram shown on the right in 
fig.~(\ref{fig_quadruplebox}).
Although there can be more than one chain diagram associated to a given Feynman graph, this non-uniqueness does not affect our algorithm.
It is sufficient to pick one chain diagram out of all possible chain diagrams.
The introduction of chains can be viewed as a convenient tool to group propagators together. 
The algorithm for the contour deformation will treat all propagators within a given chain in the same way.
The most general chain diagrams for two-loops and three-loops are shown in fig.~(\ref{fig_chain}).
The two-loop chain diagram consists of three chains, the three-loop chain diagram consists of six chains.

We define a cycle to be a closed circuit in the diagram. We can denote a cycle by specifying the chains which belong to the cycle.
In the two-loop diagram of fig.~(\ref{fig_chain}) there are three different cycles, given by
\bq
 C^{(12)}, C^{(13)}, C^{(23)}.
\eq
Here we used the notation that $C^{(ij)}$ denotes the cycle consisting of the chains $C^{(i)}$ and $C^{(j)}$.
Similar, there are seven different cycles for the three-loop chain diagram of fig.~(\ref{fig_chain}), given by
\bq
 C^{(123)}, C^{(146)}, C^{(245)}, C^{(356)}, C^{(1256)}, C^{(1345)}, C^{(2346)}.
\eq
We used the notation that $C^{(ijk)}$ and $C^{(ijkl)}$ denote cycles consisting of the chains $C^{(i)}, C^{(j)}, C^{(k)}$ and
$C^{(i)}, C^{(j)}, C^{(k)}, C^{(l)}$, respectively.
We emphasis the difference between the number of cycles of a chain diagram and the number of independent loop momenta of a chain diagram,
the former being always greater or equal to the latter one.
The loop number corresponds to the number of independent cycles and not to the number of cycles.
This distinction is relevant for the work presented here: 
The algorithm for the contour deformation in the multi-loop case will be based on the set of all cycles.
We will also show that the restriction to an independent set of cycles will in general not be sufficient.

% ----------------------------------------------------------------------------------
\section{The algorithm for the contour deformation}
\label{sect:algorithm}

We consider a $l$-loop integral with $n$ internal propagators corresponding to $c$ chains.
We have
\bq
 l \le c \le n.
\eq
Internal momenta belonging to the same chain differ only by a linear combination of the external momenta.
Without loss of generality we can order the propagators $D_i$ and the momenta $q_i$, such that
\begin{enumerate}
\item the first $l$ momenta $q_i$ coincide with the $l$ independent loop momenta $k_i$:
\bq
 q_i & = & k_i, 
 \;\;\;\;\;\; 
 1 \le i \le l.
\eq
\item the first $c$ momenta $q_i$ correspond to different chains.
\end{enumerate}
We can associate a loop momentum to each chain. With the ordering as above the loop momenta $k_1$, ..., $k_l$ are associated to the first $l$ chains.
For the remaining $(c-l)$ chains we set
\bq
\label{chain_loop_momenta}
 k_i & = & \sum\limits_{j=1}^l \rho_{ij} k_j,
 \;\;\;\;\;\; 
 l < i \le c.
\eq
The contour deformation for a $l$-loop integral is defined by the deformation of the $l$ independent loop momenta
\bq
 k_r & = & 
 \tilde{k}_r + i \lambda \kappa_r\left(\tilde{k}_1,...,\tilde{k}_l\right),
 \;\;\;\;\;\;
 1 \le r \le l,
\eq 
where all $\tilde{k}_r^{\mu}$ is real. 
Note that $\kappa_r$ depends in general on all $\tilde{k}_1$ ,..., $\tilde{k}_l$ and not just $\tilde{k}_r$
After this deformation our integral equals
\bq
I  & = &
 \int \prod\limits_{r=1}^{l} \frac{d^4\tilde{k}_r}{(2\pi)^4}
 \;
 \left|\frac{\partial k_i^{\mu}}{\partial \tilde{k}_j^{\nu}}\right|
 \;
 R(k_1,...,k_l)
 \prod\limits_{j=1}^{n} D_j.
\eq
The Jacobian 
\bq
 \left|\frac{\partial k_i^{\mu}}{\partial \tilde{k}_j^{\nu}}\right|
\eq
is the determinant of a $(4l \times 4l)$ matrix and can be computed numerically.
The vectors $\kappa_1$, ..., $\kappa_l$ provide the directions for the deformation, while the parameter $\lambda$ determines the scale
of the deformation.

\subsection{The direction of the deformation}

Let $S$ be the set of all cycles. We denote by $S_r$ the sub-set of the cycles which contain the chain $C^{(r)}$.
\begin{figure}
\begin{center}
\includegraphics[width=0.9\textwidth]{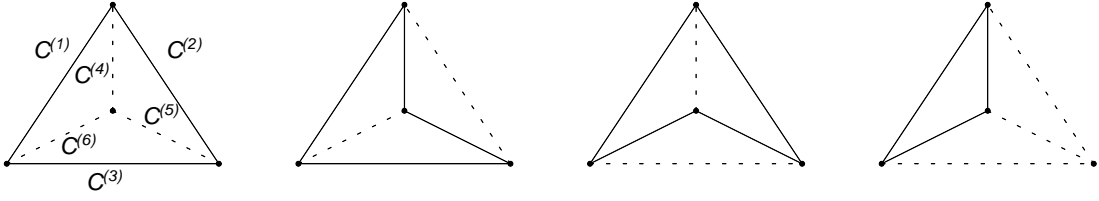}
\caption{\label{fig_chain-red} 
The four cycles $C^{(123)}$, $C^{(1345)}$, $C^{(1256)}$ and $C^{(146)}$ of the three-loop chain diagram containing the chain $C^{(1)}$.
}
\end{center}
\end{figure} 
In the example of the three-loop chain diagram we obtain for $r=1$ the set of the four cycles
\bq
 S_1 & = & 
 \left\{
 C^{(123)}, C^{(146)}, C^{(1256)}, C^{(1345)}
 \right\}
\eq
This is illustrated in fig.~(\ref{fig_chain-red}).
The set $S_r$ will be used in the construction of $\kappa_r$: The deformation vector $\kappa_r$ is given as a sum
of deformation vectors $\kappa^{(\alpha)}$, where each $\kappa^{(\alpha)}$ corresponds to one cycle in $S_r$:
\bq
\label{sum_over_cycles}
 \kappa_r & = &
 \sum\limits_{\alpha \in S_r} \kappa^{(\alpha)}
\eq
$\kappa^{(\alpha)}$ corresponds to an individual cycle. The deformation vector $\kappa^{(\alpha)}$ can be obtained with one-loop methodes by treating
all momenta not belonging to this cycle as external.
The construction of the deformation vector in the one-loop case is summarised in appendix~\ref{sect:one_loop}.
Eq.~(\ref{sum_over_cycles}) reduces therefore the problem of the construction of the deformation direction in the multi-loop case 
to the simpler problem of the construction of the deformation in the one-loop case.
Eq.~(\ref{sum_over_cycles}) is the main result of this paper.

Let us discuss the most important example:
For the two-loop chain diagram of fig.~(\ref{fig_chain}) we have
\bq
 \kappa_1 & = & \kappa^{(12)} + \kappa^{(13)},
 \nonumber \\
 \kappa_2 & = & \kappa^{(12)} + \kappa^{(23)},
\eq
while for the three-loop chain diagram of fig.~(\ref{fig_chain}) we have
\bq
 \kappa_1 & = & \kappa^{(123)} + \kappa^{(146)} + \kappa^{(1256)} + \kappa^{(1345)},
 \nonumber \\
 \kappa_2 & = & \kappa^{(123)} + \kappa^{(245)} + \kappa^{(1256)} + \kappa^{(2346)},
 \nonumber \\
 \kappa_3 & = & \kappa^{(123)} + \kappa^{(356)} + \kappa^{(1345)} + \kappa^{(2346)}.
\eq
The reader might worry that the definition of the contour deformation depends on the choice of the independent loop momenta and/or singles
out particular cycles (for example the cycle $C^{(12)}$ appearing twice in the two-loop case, while the cycles $C^{(13)}$ and $C^{(23)}$ only appear once).
This is not the case.
If we set in the analogy with eq.~(\ref{chain_loop_momenta})
\bq
 \kappa_i & = & \sum\limits_{j=1}^l \rho_{ij} \kappa_j,
 \;\;\;\;\;\; 
 l < i \le c,
\eq
we have in the two-loop case
\bq
\label{example_deformation_full_two_loop}
 \kappa_1 & = & \kappa^{(12)} + \kappa^{(13)},
 \nonumber \\
 \kappa_2 & = & \kappa^{(12)} + \kappa^{(23)},
 \nonumber \\
 \kappa_3 & = & \kappa^{(13)} - \kappa^{(23)},
\eq
and in the three-loop case
\bq
 \kappa_1 & = & \kappa^{(123)} + \kappa^{(146)} + \kappa^{(1256)} + \kappa^{(1345)},
 \nonumber \\
 \kappa_2 & = & \kappa^{(123)} + \kappa^{(245)} + \kappa^{(1256)} + \kappa^{(2346)},
 \nonumber \\
 \kappa_3 & = & \kappa^{(123)} + \kappa^{(356)} + \kappa^{(1345)} + \kappa^{(2346)},
 \nonumber \\
 \kappa_4 & = & \kappa^{(146)} - \kappa^{(245)} + \kappa^{(1345)} - \kappa^{(2346)},
 \nonumber \\
 \kappa_5 & = & \kappa^{(245)} - \kappa^{(356)} + \kappa^{(1256)} - \kappa^{(1345)},
 \nonumber \\
 \kappa_6 & = & \kappa^{(356)} - \kappa^{(146)} + \kappa^{(2346)} - \kappa^{(1256)}.
\eq
The deformation in each chain is given by the signed sum of deformation vectors over all cycles which contain
the given chain. The signs follow from the orientation of the momentum flow.

We would like to comment on the sum over the cycles $S_r$ in eq.~(\ref{sum_over_cycles}).
As an example we discuss the two-loop case as in eq.~(\ref{example_deformation_full_two_loop}).
Suppose that the momentum flowing through one of the propagators of the chain $C^{(3)}$ becomes soft.
Then the deformation vector corresponding to any cycle containing the chain $C^{(3)}$ goes to zero.
In the example at hand this implies
\bq
 \kappa^{(13)} \rightarrow 0,
 & &
 \kappa^{(23)} \rightarrow 0,
\eq
and as a consequence
\bq 
 \kappa_3 & \rightarrow & 0,
\eq
as it should be.
Suppose now that $\kappa^{(12)}$ would not be present in eq.~(\ref{example_deformation_full_two_loop}). 
Then also $\kappa_1$ and $\kappa_2$ would go to zero.
However this is not correct. The deformation for the chains $C^{(1)}$ and $C^{(2)}$ is not pinched and $\kappa^{(12)}$ provides the correct
deformation in this case.
This shows that deformation corresponding to all cycles have to be taken into account.

Let us consider a deformation vector $\kappa^{(\alpha)}$ corresponding to an individual cycle $\alpha$. 
This deformation vector is constructed in such a way that whenever one propagator of the cycle $\alpha$ goes on-shell ($q_i^2-m_i^2 \rightarrow 0$) 
we have $2 q_i \cdot \kappa^{(\alpha)} \ge 0$.
Now pick out any propagator $i$ of the original Feynman graph. 
This propagator belongs to several cycles. Let us denote this set by $S_r$.
The deformation vector $\kappa_r$ for the propagator is the (signed) sum over the deformation vectors $\kappa^{(\alpha)}$
corresponding to the cycles contained in $S_r$.
(The sign takes care of the relative orientation.)
For simplicity of the argument let us assume that all signs are positive.
For each $\alpha$ we have then $2 q_i \cdot \kappa^{(\alpha)} \ge 0$ whenever $q_i^2-m_i^2 \rightarrow 0$.
It follows that this inequality also holds for the sum of all $\kappa^{(\alpha)}$:
We have $2 q_i \cdot \kappa_r \ge 0$ whenever $q_i^2-m_i^2 \rightarrow 0$.

\subsection{The scaling parameter}

The deformation vectors $\kappa_r$ defined above provide the right direction for the contour deformation.
The deformation is correct, if the deformation vectors would be infinitesimal.
However, in practice we would like them to be of finite length and in fact to take them as large as possible.
We can think of starting from an infinitesimal length and scale the deformation vectors to the largest allowed length.
The largest allowed length is determined by the nearest pole of the propagators in this direction.
The scaling parameter takes care of this.
The definition of the scaling parameter is very similar to the one-loop case.
The scaling parameter $\lambda$ is given by
\bq
\lambda&=& \min\left[1,\lambda_{1},...,\lambda_{n},\lambda_{\mathrm{UV}}^1,...,\lambda_{\mathrm{UV}}^{|S|}\right],
\eq
with
\bq
\label{def_lambda_j}
\lambda_{j}^{2}&=&\left\{\begin{array}{rcl}
Y_{j}/4&:& 2X_{j}\ < \ Y_{j} \\
X_{j}-Y_{j}/4&:&0\ <\ Y_{j}\ <\ 2X_{j} \\
X_{j}-Y_{j}/2&:& Y_{j}\ <\ 0\end{array}\right.
\eq
and
\bq
\label{def_Xj_Yj}
 X_{j} 
 & = &
 \left(\frac{\kappa_{j}\cdot \tilde{q}_{j}}{\kappa_{j}^{2}}\right)^{2}
 \quad\mbox{and}\quad 
 Y_{j}\ = \ \frac{\tilde{q}_{j}^{2}-m_{j}^{2}}{\kappa_{j}^{2}}.
\eq
Here we used the notation that the momentum flowing through the $j$-th propagator is given by
\bq
 \tilde{q}_j + i \kappa_j,
\eq
where both $\tilde{q}_j$ and $\kappa_j$ have real entries.
The values $\lambda_1$, ..., $\lambda_n$ ensure that we do not hit accidently one of the poles of the $n$ propagators.
On the other hand, the values $\lambda_{\mathrm{UV}}^1$, ..., $\lambda_{\mathrm{UV}}^{|S|}$ ensure that we do not hit accidently 
one of the (harmless) poles of the function
$R(k_1,...,k_l)$. We assume that we have $|S|$ ultraviolet propagators, one for each cycle.
We define the parameters $\lambda_{\mathrm{UV}}^j$ by
\bq
\lambda_{\mathrm{UV}}^j
 & = &
\left\{\begin{array}{rcl}
1 & : & 4 \kappa_j \cdot \tilde{\bar{q}}_j \  > \ \operatorname{Im}(\mu_{\mathrm{UV}}^{2}) \\
\frac{\operatorname{Im}(\mu_{\mathrm{UV}}^{2})}{4 \kappa_j \cdot \tilde{\bar{q}}_j} & : &  4 \kappa_j \cdot \tilde{\bar{q}}_j \ \leq \ \operatorname{Im}(\mu_{\mathrm{UV}}^{2})
\end{array}\right.
\eq
Also here we used the notation that the momentum flowing through the $j$-th ultraviolet propagator is given by
\bq
 \tilde{\bar{q}}_j + i \kappa_j,
\eq
where both $\tilde{\bar{q}}_j$ and $\kappa_j$ have real entries.

% ----------------------------------------------------------------------------------
\section{Checks and examples}
\label{sect:checks}

\begin{figure}
\begin{center}
\includegraphics[viewport=185 565 285 630]{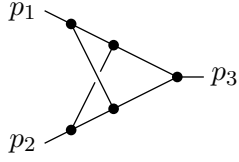}
\end{center}
\caption{
The non-planar two-loop three-point function. 
}
\label{fig_crossedtraingle}
\end{figure}

\begin{figure}
\begin{center}
\includegraphics[scale=0.9]{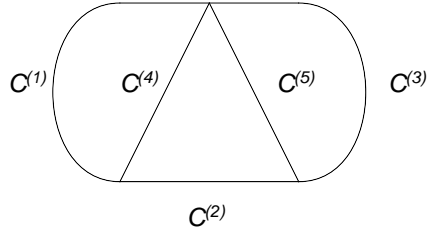}
\caption{\label{fig_degenerate_chain_diagram} 
The degenerate chain diagram of an three-loop ladder diagram.
}
\end{center}
\end{figure}

In this section we test the algorithm for the contour deformation for multi-loop integrals by calculating several scalar Feynman integral
and by comparing our results with known analytical results wherever they are available.
We consider scalar multi-loop integrals, where all external particles are off-shell. These integrals are infrared finite.
In addition we require that the integrals are also ultraviolet finite. In particular this excludes diagrams with one-loop self-energy type sub-diagrams.
Since the examples which we consider are ultraviolet and infrared finite we do not need any subtraction terms.
The normalisation of the integrals is in accordance with eq.~(\ref{eq_basic_integral})
\bq
I  & = &
 \int \prod\limits_{r=1}^{l} \frac{d^4k_r}{(2\pi)^4}
 \;
 \prod\limits_{j=1}^{n} D_j,
\eq
where $D_j$ denotes the propagators as in eq.~(\ref{def_propagator}).
In the literature there are simple analytical results for the two-, three- and four-point ladder diagrams with off-shell external momenta
and massless internal lines \cite{Usyukina:1993ch}.
\begin{figure}
\begin{center}
\includegraphics[width=0.75\textwidth]{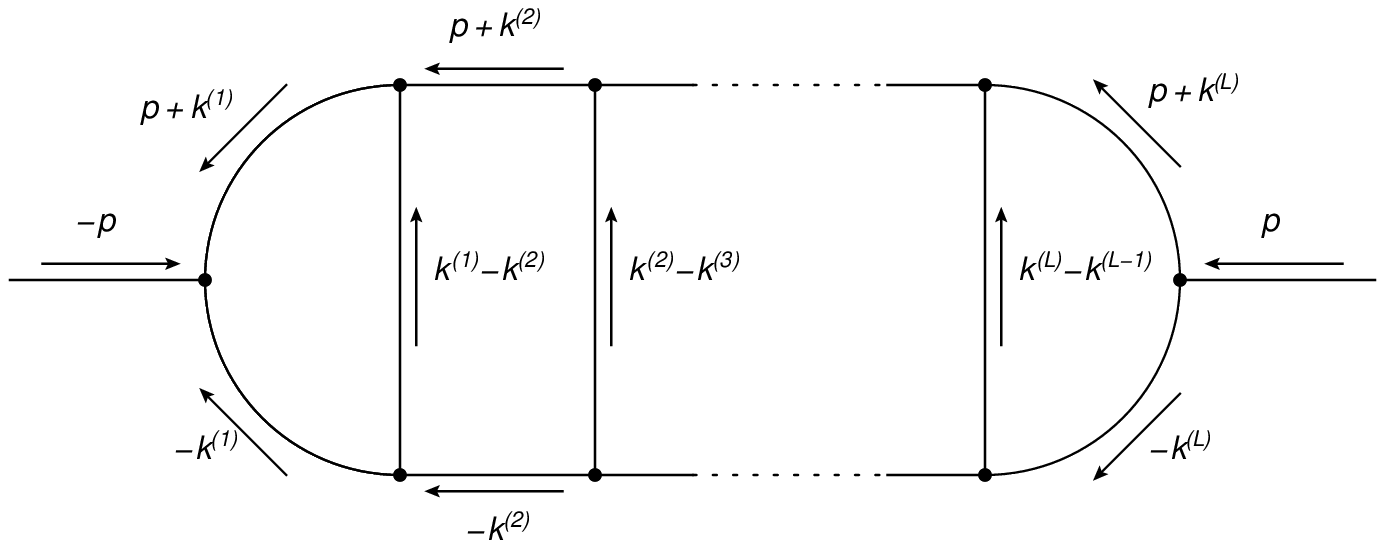}
\\
\vspace*{15mm}
\includegraphics[width=0.75\textwidth]{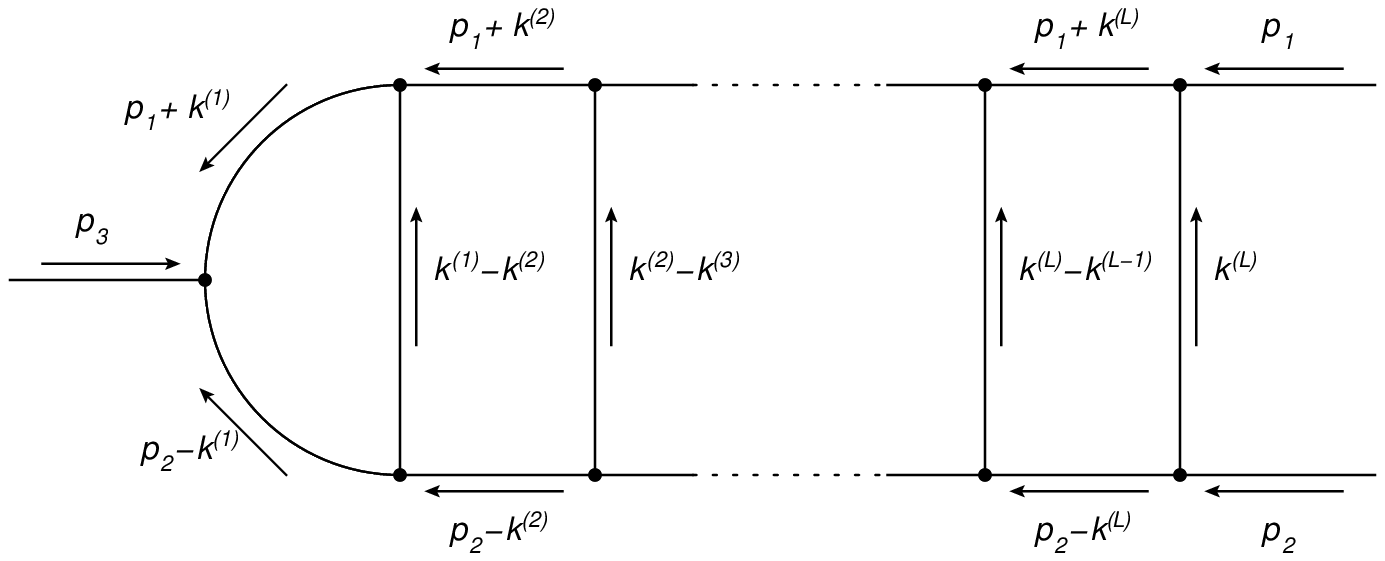}
\\
\vspace*{15mm}
\includegraphics[width=0.75\textwidth]{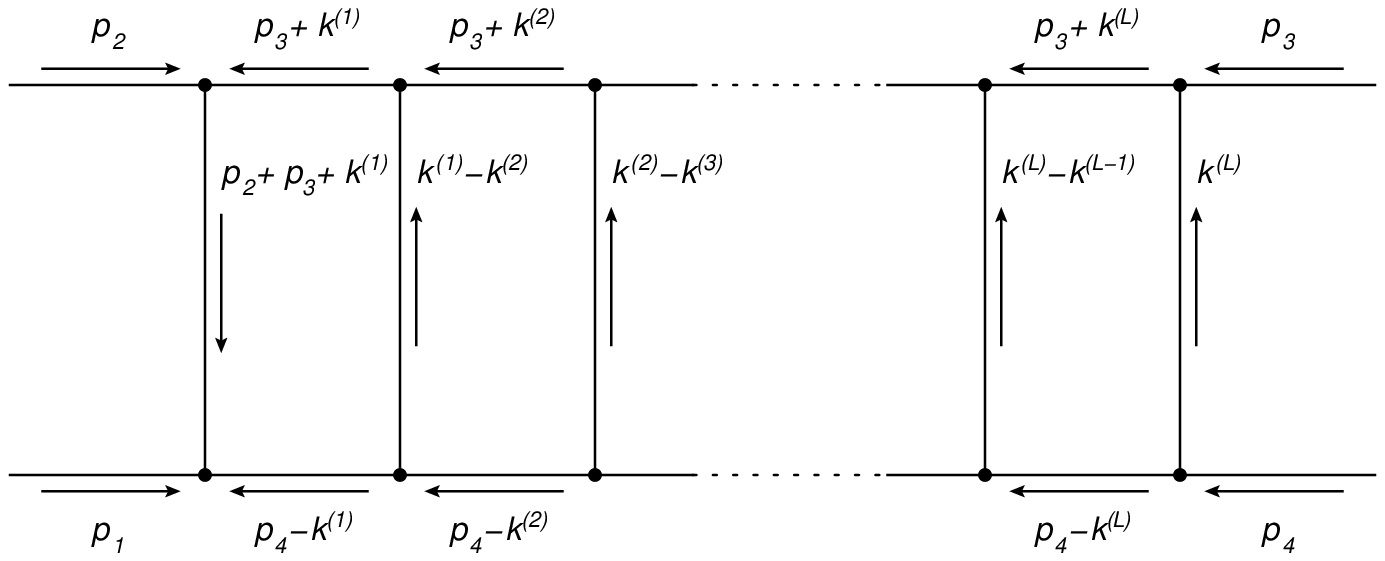}
\caption{\label{fig_ladder_type} 
The two-, three- and four-point ladder diagrams.
}
\end{center}
\end{figure} 
These diagrams are shown in fig.~(\ref{fig_ladder_type}).
We denote the $l$-loop two-point ladder integral by $B^{(l)}(p^{2})$,
the $l$-loop three-point ladder integral by $C^{(l)}(p_{1}^{2},p_{2}^{2},p_{3}^{2})$
and the $l$-loop four-point ladder integral by $D^{(l)}(p_{1}^{2},p_{2}^{2},p_{3}^{2},p_{4}^{2},s,t)$.
We briefly recall the analytical results for these integrals.
We have
\bq
\label{def_ladder_integrals}
 B^{(l)}(p^{2})
 & = &
 \left(\frac{i}{16 \pi^2 p^{2}}\right)^{l}p^{2} \frac{(2l)!}{(l!)^{2}}\zeta_{2l-1},
 \nonumber \\
 C^{(l)}(p_{1}^{2},p_{2}^{2},p_{3}^{2}) 
 & = & 
 \left(\frac{i}{16 \pi^2 p_{3}^{2}}\right)^{l}\Phi^{(l)}(x,y),
 \nonumber \\
 D^{(l)}(p_{1}^{2},p_{2}^{2},p_{3}^{2},p_{4}^{2},s,t) 
 & = & 
 \left(\frac{i}{16 \pi^2 s}\right)^{l}\frac{1}{t}\Phi^{(l)}(X,Y).
\eq
The function $\Phi$ can be expressed in terms of polylogarithms 
\bq
\Phi^{(l)}(x,y)
 & = &
 -
 \frac{1}{l!\lambda(x,y)}\sum\limits_{j=l}^{2l}\frac{(-1)^{j}j!\ln^{2l-j}(y/x)}{(j-l)!(2l-j)!}
  \left[\operatorname{Li}_{j}\left(-\frac{1}{x \rho(x,y)}\right)-\operatorname{Li}_{j}(-y \rho(x,y))\right]. 
\;\;\;
\eq
The definitions of the variables are
\begin{eqnarray}
\begin{array}{rclcrcl}
\displaystyle x& = &\displaystyle \frac{p_{1}^{2}}{p_{3}^{2}},&\qquad &\displaystyle  y & = &\displaystyle \frac{p_{2}^{2}}{p_{3}^{2}},\\[0.4cm]
\displaystyle X& = &\displaystyle \frac{p_{1}^{2}p_{3}^{2}}{st},&\qquad &\displaystyle  Y & = &\displaystyle  \frac{p_{2}^{2}p_{4}^{2}}{st},\\[0.4cm]
s& = &\displaystyle  (p_{1}+p_{2})^{2},&\qquad& t& = &\displaystyle  (p_{2}+p_{3})^{2},\\[0.4cm]
\displaystyle \lambda(x,y)&=&\displaystyle \sqrt{(1-x-y)^{2}-4xy},&\qquad&
\displaystyle \rho(x,y)&=&\displaystyle \frac{2}{1-x-y+\lambda}.
\end{array}
\end{eqnarray}
In addition we consider the non-planar two-loop three-point function shown in fig.~(\ref{fig_crossedtraingle}), again with off-shell external momenta and
massless internal lines.
The analytical result for this integral is given by \cite{Usyukina:1994iw}
\bq
C_{\mathrm{np}}^{(2)}(p_{1}^{2},p_{2}^{2},p_{3}^{2})
 & = & 
 \left(C^{(1)}(p_{1}^{2},p_{2}^{2},p_{3}^{2})\right)^{2}.
\eq
The chain diagram of the three-loop ladder graph is degenerate and can be obtained from the general three-loop chain diagram in fig.~(\ref{fig_chain})
by pinching one of the chains $C^{(4)}$, $C^{(5)}$ or $C^{(6)}$.
The resulting chain diagram is shown in fig.~(\ref{fig_degenerate_chain_diagram}).
As a consequence the deformation vectors simplify to
\bq
 \kappa_1 & = & \kappa^{(123)} + \kappa^{(14)} + \kappa^{(125)},
 \nonumber \\
 \kappa_2 & = & \kappa^{(123)} + \kappa^{(245)} + \kappa^{(125)} + \kappa^{(234)},
 \nonumber \\
 \kappa_3 & = & \kappa^{(123)} + \kappa^{(35)} + \kappa^{(234)},
 \nonumber \\
 \kappa_4 & = & \kappa^{(14)} - \kappa^{(245)} - \kappa^{(234)},
 \nonumber \\
 \kappa_5 & = & \kappa^{(245)} - \kappa^{(35)} + \kappa^{(125)}.
\eq
We present the results of the numerical evaluation together with the analytical results 
for the two and three-loop ladder diagrams defined in eq.~(\ref{def_ladder_integrals}) and the non-planar two-loop three-point integral in table~(\ref{table_ladder_integrals}).
\begin{table}
\begin{center}
\begin{tabular}{|c|c|c|}
 \hline
 integral & numerical result & analytical result \\ 
 \hline
 $B^{(2)} [\mathrm{GeV}^{-2}]$ & $(-3.59 \pm 0.05) \cdot 10^{-8}$ & $-3.571 \cdot 10^{-8}$ \\
 $C^{(2)} [\mathrm{GeV}^{-4}]$ & $(-1.80 \pm 0.05) \cdot 10^{-11}$ & $-1.832 \cdot 10^{-11}$ \\
 $C^{(2)}_{\mathrm{np}} [\mathrm{GeV}^{-4}]$ & $(-2.93 \pm 0.04) \cdot 10^{-11}$ & $-2.904 \cdot 10^{-11}$ \\
 $D^{(2)} [\mathrm{GeV}^{-6}]$ & $(-5.88 \pm 0.07) \cdot 10^{-14}$ & $-5.897 \cdot 10^{-14}$ \\
 \hline
 $B^{(3)} [\mathrm{GeV}^{-4}]$ & $(-7.9 \pm 0.5) \cdot 10^{-14} i$ & $-8.027 \cdot 10^{-14} i$ \\
 $C^{(3)} [\mathrm{GeV}^{-6}]$ & $(-5.3 \pm 0.6) \cdot 10^{-17} i$ & $-5.389 \cdot 10^{-17} i$ \\
 $D^{(3)} [\mathrm{GeV}^{-8}]$ & $(-7.1 \pm 0.7) \cdot 10^{-19} i$ & $-6.744 \cdot 10^{-19} i$ \\
 \hline
\end{tabular}
\caption{\label{table_ladder_integrals} Results for the various two- and three-loop ladder diagrams.}
\end{center}
\end{table}
The numerical values of the external momenta are given for the two-point functions by
\bq
 p & = & \left( 90, 0, 0, 0 \right) \; \mbox{GeV}.
\eq
For the three-point functions the external momenta are given by
\bq
 p_1 & = & \left( 39.7424, -14.1093, 0.102709, 20.4908 \right) \; \mbox{GeV},
 \nonumber \\
 p_2 & = & \left( 50.2576, 14.1093, -0.102709, -20.4908 \right) \; \mbox{GeV},
 \nonumber \\
 p_3 & = & \left( -90, 0, 0, 0 \right) \; \mbox{GeV}.
\eq
For the four-point functions the external momenta are given by
\bq
 p_1 & = & \left( 19.6586, -7.15252, -0.206016, 8.96383 \right) \; \mbox{GeV},
 \nonumber \\
 p_2 & = & \left( 26.874, 7.04203, -0.0501295, -12.9055 \right) \; \mbox{GeV},
 \nonumber \\
 p_3 & = & \left( 43.4674, 0.110491, 0.256146, 3.9417 \right) \; \mbox{GeV},
 \nonumber \\
 p_4 & = & \left( -90, 0, 0, 0 \right) \; \mbox{GeV}.
\eq
In table~(\ref{table_ladder_integrals}) we observe a good agreement between the known analytical results and our numerical evaluations.
The results for the three-loop integrals have a larger Monte Carlo integration error as compared to the two-loop integrals.
This is of course expected.

In addition we include here results on the two-loop six-point functions, again with massless internal lines.
In this case no analytical results are available.
\begin{figure}
\begin{center}
\includegraphics[scale=1.4]{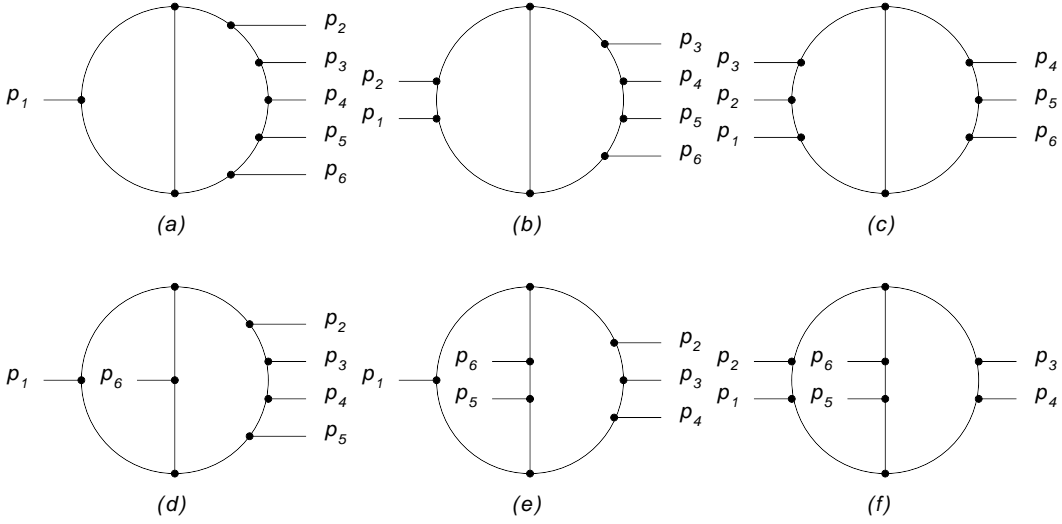}
\caption{\label{fig_2loop6point} 
Definition of the various two-loop six-point topologies.
}
\end{center}
\end{figure} 
The various two-loop six-point topologies are shown in fig.~(\ref{fig_2loop6point}).
The external momenta are defined by
\bq
\label{external_momenta_2loop_6point}
 p_1 & = & \left( 12.0588, -1.00017, -2.55373, 2.65288 \right) \; \mbox{GeV},
 \nonumber \\
 p_2 & = & \left( 18.6089, -8.9195, 6.43508, 9.61832 \right) \; \mbox{GeV},
 \nonumber \\
 p_3 & = & \left( 13.8389, -4.73227,-6.55009, -1.55854 \right) \; \mbox{GeV},
 \nonumber \\
 p_4 & = & \left( 25.5377, 20.892, 4.32472, -8.89684 \right) \; \mbox{GeV},
 \nonumber \\
 p_5 & = & \left( 19.9556, -6.24009, -1.65597, -1.81582 \right) \; \mbox{GeV},
 \nonumber \\
 p_6 & = & \left( -90, 0, 0, 0 \right) \; \mbox{GeV},
 \nonumber \\
\eq
Our results for the real part can be found in table~(\ref{table_twoloop_sixpoint}).
\begin{table}
\begin{center}
\begin{tabular}{|c|c|}
 \hline
 topology & our result \\ 
 \hline
 (a) $[\mathrm{GeV}^{-10}]$ & $(-8.66 \pm 0.08) \cdot 10^{-19}$ \\
 (b) $[\mathrm{GeV}^{-10}]$ & $(-1.17 \pm 0.02) \cdot 10^{-18}$ \\
 (c) $[\mathrm{GeV}^{-10}]$ & $(-7.75 \pm 0.13) \cdot 10^{-19}$ \\
 (d) $[\mathrm{GeV}^{-10}]$ & $(-1.91 \pm 0.02) \cdot 10^{-19}$ \\
 (e) $[\mathrm{GeV}^{-10}]$ & $(-4.64 \pm 0.08) \cdot 10^{-19}$ \\
 (f) $[\mathrm{GeV}^{-10}]$ & $(-1.03 \pm 0.02) \cdot 10^{-18}$ \\
 \hline
\end{tabular}
\caption{\label{table_twoloop_sixpoint} Results for the real part of the various two-loop six-point integrals.}
\end{center}
\end{table}
Table~(\ref{table_twoloop_sixpoint}) demonstrates that two-loop six-point functions can be computed easily.
The numerical computation of the imaginary part gives results compatible with zero within an error which is of the order of the error of the real part.
This is of course expected, since the kinematical configuration in eq.~(\ref{external_momenta_2loop_6point}) corresponds to a $1\rightarrow 5$ process
and all invariants entering the graph polynomial are positive.

In addition we have checked that the algorithm works as expected by adding masses to the internal propagators.

% ----------------------------------------------------------------------------------
\section{Conclusions}
\label{sect:conclusions}

In this paper we discussed an algorithm for the contour deformation in loop momentum space for multi-loop integrals.
The method can be applied to finite multi-loop integrals and to divergent multi-loop integrals with suitable subtraction terms.
The problem is reduced to the one-loop case through a sum over all possible cycles.
We demonstrated on non-trivial two- and three-loop examples that the method works correctly.
The techniques developed in this paper will be useful for a numerical approach towards NNLO corrections and beyond.

\subsection*{Acknowledgements}

This work is supported by the U.S. Department of Energy, Basic Energy Sciences, Office of Science,
under contract DE-AC02-06CH11357.

% ----------------------------------------------------------------------------------

\begin{appendix}

\section{The contour deformation for an individual cycle}
\label{sect:one_loop}

In this appendix we discuss the construction of the deformation vector for an individual cycle $C^{(\alpha)}$.
We assume that the cycle consists of $n^{(\alpha)}$ propagators.
For a cycle we construct the vector $\kappa^{(\alpha)}$, which gives the deformation direction.
For the construction of the deformation vector $\kappa^{(\alpha)}$ we work in a specific Lorentz frame, which we take to be the
centre-of-mass frame defined by the external momenta of the multi-loop integral under considerations (or more generally the external
momenta of the scattering process under consideration).
We denote the centre-of-mass energy by $\sqrt{\hat{s}}$.
This is a characteristic scale of the problem.

An individual cycle can be viewed as a one-loop diagram, where all momenta not belonging to the cycle are treated
as external momenta.
In the following we will drop the super-script $\alpha$ and adopt the notation of the one-loop case.
We consider the pole structure
\bq
 \prod\limits_{j=1}^n \frac{1}{\left(k-q_j\right)^2 - m_j^2},
\eq
with $n$ propagators. The loop momentum is denoted by $k$, and the origins of the $n$ cones are given by the four-vectors $q_1$, ..., $q_n$.
The masses are given by $m_1$, ..., $m_n$.
The method presented here is a minor modification of the one given in ref.~\cite{Becker:2012nk}.
The modifications concern the definition of the vectors in eq.~(\ref{def_kplusminus}) below.
In ref.~\cite{Becker:2012nk} we considered the physical situation of a $2 \rightarrow (n-2)$ process.
In this situation two external particles (the two initial particles) have a negative energy component. 
In the general situation we have to consider the case of an arbitrary number of external particles with negative energy component.
In order to simplify the notation we write in the following for the real loop momentum $k$ instead of $\tilde{k}$.
The various helper functions occuring in the definition of the deformation vector and the default values of the technical parameters are summarised
at the end of this appendix.

\subsection{The deformation vector}

The deformation vector $\kappa$ is constructed as a sum of two terms and a scaling parameter
\bq
\kappa & = & \lambda^{\mathrm{cycle}} \left( \kappa_{\mathrm{int}}+\kappa_{\mathrm{ext}} \right).
\eq
$\kappa_{\mathrm{ext}}$ is given by
\bq
\kappa_{\mathrm{ext}}^{\mu}(k)&=&g_{\mu\nu}\left(c_{+}k^{\nu}_{+}+c_{-}k^{\nu}_{-}\right) 
\eq
with
\bq
\label{def_kplusminus}
k_{\pm}& = & k-P_{\pm}.
\eq
and $g_{\mu\nu}=\mbox{diag}(1,-1,-1,-1)$ the metric tensor.
The four-vectors $P_{\pm}$ are chosen such that all points $q_1$, ..., $q_n$ are in the forward light-cone of $P_+$ and in the backward light-cone of $P_-$.
We construct $P_+$ iteratively through the following algorithm:
We start with the list $(q_1,...,q_n)$ and go through the following steps:
\begin{enumerate}
\item Remove all four-vectors, which are in the forward light-cone of some other four-vector.
(After this step all four-vectors are separated by space-like distances.)
\item Find the pair $(q_i,q_j)$ 
with the smallest space-like separation, i.e. with the smallest value of $-(q_i-q_j)^2$.
\item Replace $q_i$ and $q_j$ by the combined four-vector
\bq
 Z_+\left(q_i+q_j, q_i-q_j\right),
 & &
Z_+^{\mu}(x,y) = \frac{1}{2}\left(x^{\mu} + \frac{y_{\nu}}{|\vec{y}|}\left(g^{0\mu}y^{\nu}-g^{0\nu}y^{\mu}\right)\right).
\eq
\item Go back to step 1 until only one entry is left. Set the final entry equal to $P_+$.
\end{enumerate}
The construction of $P_-$ proceeds analogously:
\begin{enumerate}
\item Remove all four-vectors, which are in the backward light-cone of some other four-vector.
(Again, after this step all four-vectors are separated by space-like distances.)
\item Find the pair $(q_i,q_j)$ 
with the smallest space-like separation, i.e. with the smallest value of $-(q_i-q_j)^2$.
\item Replace $q_i$ and $q_j$ by the combined four-vector
\bq
 Z_-\left(q_i+q_j, q_i-q_j\right),
 & &
Z_-^{\mu}(x,y) = \frac{1}{2}\left(x^{\mu} - \frac{y_{\nu}}{|\vec{y}|}\left(g^{0\mu}y^{\nu}-g^{0\nu}y^{\mu}\right)\right).
\eq
\item Go back to step 1 until only one entry is left. Set the final entry equal to $P_-$.
\end{enumerate}
This algorithm constructs $P_\pm$ independent of the order of the list $(q_1,...,q_n)$.
The four-vector $(P_--P_+)$ is by construction always time-like.
We define
\bq
\label{def_mu_P}
 \mu_P & = & \sqrt{\left(P_- - P_+\right)^2}.
\eq
$\mu_P$ defines a characteristic scale associated to the cycle under consideration.
The coefficients $c_\pm$ are defined as
\bq
c_{\pm} & = & \prod\limits_{i=1}^{n}h_{\delta\mp}\left(k_{i},m_{i}^{2},M_3^2\right),
\eq
A typical value for the parameter $M_3^2$ is given by 
\bq 
 M_3 & = & 0.035 \; \max\left(\mu_P, \sqrt{\hat{s}} \right),
\eq
where $\sqrt{\hat{s}}$ is the centre-of-mass energy
defined by the external momenta of the multi-loop integral and
$\mu_P$ has been defined in eq.~(\ref{def_mu_P}).

$\kappa_{\mathrm{int}}$ is given by
\bq
\label{def_massive_interior}
\kappa_{\mathrm{int}}^{\mu}&=&-\sum\limits_{i=1}^{n}c_{i}k_{i}^{\mu}-\sum\limits_{\substack{i,j=1\\i<j}}^{n}c_{ij}k^{\mu}_{ij}
+ \kappa_{\mathrm{soft}}^{\mu}
\eq
with
\bq
k_{i}&=&k-q_{i} \qquad\mbox{and}\qquad k_{ij}\ = \ k-v_{ij}.
\eq
The four-vectors $v_{ij}$ are defined by
\bq
 v_{ij} & = & \frac{1}{2} \left( q_i + q_j - \frac{(m_i-m_j)}{\sqrt{(q_i-q_j)^2}} (q_i-q_j) \right).
\eq
The coefficients are given by
\bq
c_{i} & = & g\left(k_{\mathrm{centre}},\gamma_1,M_2^2\right) \prod\limits_{l=1}^{n}d_{i,l}
 \qquad\mbox{and}\qquad
c_{ij}\ =\ g\left(k_{\mathrm{centre}},\gamma_1,M_2^2\right) \prod\limits_{l=1}^{n}d_{ij,l},
\eq
with 
\bq
 k_{\mathrm{centre}} & = & \frac{1}{2} \left( k_+ + k_- \right).
\eq
Typical values for the parameters $\gamma_1$ and $M_2^2$ are $\gamma_1=0.7$ and 
\bq
 M_2 & = & 0.7 \; \max\left( \mu_P, \sqrt{\hat{s}} \right).
\eq
The factors $d_{i,l}$ are defined by
\bq
\label{def_dil}
d_{i,l} & = &
 \left\{\begin{array}{rcl} 
  1 & : & l=i,\ m_{l}=0 \\
 [0.3cm]
 h_{\delta+}(k_{l},m_{l}^{2},M_1^2) & : & (q_{i}-q_{l})^{2}=0,\ q_{i}^{0} < q_{l}^{0},\ m_{l}=0 \\
 [0.3cm]
 h_{\delta-}(k_{l},m_{l}^{2},M_1^2) & : & (q_{i}-q_{l})^{2}=0,\ q_{i}^{0} > q_{l}^{0},\ m_{l}=0 \\
 [0.3cm]
 \max\left[h_{\delta}(k_{l},m_{l}^{2},M_1^2),h_{\theta}(-2k_{l} \cdot k_{i},M_1^2)\right] &:&\mbox{otherwise}   
 \end{array} \right.
 \nonumber \\
\eq
The factors $d_{ij,l}$ are defined by
\bq
\label{def_dijl}
d_{ij,l} & = & 
 h_{\theta}(z_{ij},M_1^2) \ \max\left[h_{\delta}(k_{l},m_{l}^{2},M_1^2),h_{\theta}(-2k_{l} \cdot k_{ij},M_1^2)\right],
 \nonumber \\
 z_{ij} & = & \left(q_i-q_j\right)^2 - \left(m_i+m_j\right)^2.
\eq   
A typical value for the parameters $M_1$ is given by 
\bq
 M_1 & = & 0.035 \sqrt{\hat{s}}.
\eq
We further have
\bq
 \kappa_{\mathrm{soft}}
 & = &
 \sum\limits_{a} c_a \kappa_a,
\eq
where the index $a$ sums over some pre-defined directions.
We take the pre-defined directions to be the four Cartesian coordinate directions:
\bq
 \kappa_0 = E_{\mathrm{soft}} \left(1,0,0,0\right), & & \kappa_1 = E_{\mathrm{soft}} \left( 0,1,0,0\right),
 \nonumber \\
 \kappa_2 = E_{\mathrm{soft}} \left(0,0,1,0\right), & & \kappa_3 = E_{\mathrm{soft}} \left( 0,0,0,1\right),
\eq
with $E_{\mathrm{soft}}$ being an energy scale much smaller then the centre-of-mass scale. A typical value is $E_{\mathrm{soft}}= 0.03 \sqrt{\hat{s}}$.
The coefficients $c^a$ are given by
\bq
 c_a & = & 
 g\left(k_{\mathrm{centre}},\gamma_1,M_2^2\right) \left( \prod\limits_{l=1}^{n} d^+_{a,l} - \prod\limits_{l=1}^{n} d^-_{a,l} \right)
\eq
with 
\bq
\label{def_d_plus_minus}
 d^+_{a,l} & = &
 \max\left[h_{\delta}(k_{l},m_{l}^{2},\gamma_2 M_1^2),h_{\theta}(2k_{l} \cdot \kappa^a,\gamma_2 M_1^2)\right],
 \nonumber \\
 d^-_{a,l} & = &
 \max\left[h_{\delta}(k_{l},m_{l}^{2},\gamma_2 M_1^2),h_{\theta}(-2k_{l} \cdot \kappa^a,\gamma_2 M_1^2)\right].
\eq
A typical value for $\gamma_2$ is $\gamma_2=0.008$.

It remains to define the scaling parameter $\lambda^{\mathrm{cycle}}$ for the cycle.
We set 
\bq
 \kappa_0 & = & \kappa_{\mathrm{int}} + \kappa_{\mathrm{ext}}.
\eq
The scaling parameter $\lambda^{\mathrm{cycle}}$ is given by
\bq
\lambda^{\mathrm{cycle}}&=& \min\left[1,\lambda_{1},...,\lambda_{n},\lambda_{\mathrm{coll}}\right],
\eq
with
\bq
\lambda_{j}^{2}&=&\left\{\begin{array}{rcl}
Y_{j}/4&:& 2X_{j}\ < \ Y_{j} \\
X_{j}-Y_{j}/4&:&0\ <\ Y_{j}\ <\ 2X_{j} \\
X_{j}-Y_{j}/2&:& Y_{j}\ <\ 0\end{array}\right.
\eq
and
\bq
 X_{j} 
 & = &
 \left(\frac{\kappa_{0}\cdot k_{j}}{\kappa_{0}^{2}}\right)^{2}
 \quad\mbox{and}\quad 
 Y_{j}\ = \ \frac{k_{j}^{2}-m_{j}^{2}}{\kappa_{0}^{2}}.
\eq
Next we define a value $\lambda_{\mathrm{coll}}$ by
\bq
\lambda_{\mathrm{coll}}&=&\frac{1}{4N},
\eq
with
\bq
N&=&\sum\limits_{i=1}^{n}c_{i}+\sum\limits_{\substack{i,j=1\\i<j}}^{n}c_{ij}
 + \sum\limits_a \left| c_a \right|.
\eq

\subsection{Helper functions}

We list here the various helper functions appearing in the definition of the contour deformation.
The functions $h_{\delta+}$ and $h_{\delta-}$ are defined by
\bq
\label{dd3}
h_{\delta\pm}(k,m^{2},M_1^2)&=&\frac{\left(\pm k^{0}-\sqrt{\vec{k}^{2}+m^{2}}\right)^{2}}{\left(\pm k^{0}-\sqrt{\vec{k}^{2}+m^{2}}\right)^{2}+M_{1}^{2}}.
\eq
The function $h_{\delta}$ is given by
\bq
h_{\delta}(k,m^{2},M_1^2)&=&  
\frac{\left(|k^{0}|-\sqrt{\vec{k}^{2}+m^{2}}\right)^{2}}{\left(|k^{0}|-\sqrt{\vec{k}^{2}+m^{2}}\right)^{2}+M_{1}^{2}}.
\eq
The function $h_{\theta}(t,M_1^2)$ is defined by
\bq
 h_{\theta}\left(t,M_1^2\right)
 & = & 
 \frac{t}{t + M_1^2} \theta\left( t \right),
\eq
where $\theta(t)$ is the Heaviside step function. 
The function $g$ is given by
\bq
g(k,\gamma_1,M_2^2)&=&\frac{\gamma_1 M_2^{2}}{k\circ k+M_2^{2}}.
\eq
The operation ``$\circ$'' denotes the Euclidean scalar product of two four-vectors. 

\end{appendix}

% ----------------------------------------------------------------------------------
% references
\bibliography{/home/stefanw/notes/biblio}
\bibliographystyle{/home/stefanw/latex-style/h-physrev5}

\end{document}